\documentclass{bioinfo}
\copyrightyear{2015}
\pubyear{2015}

\usepackage{amsmath}
\usepackage[ruled,vlined]{algorithm2e}
\usepackage{natbib}

\begin{document}
\firstpage{1}

\title[Large-scale Genotype Query]{BGT: efficient and flexible genotype query across many samples}

\author[Li]{Heng Li}

\address{Broad Institute, 75 Ames Street, Cambridge, MA 02142, USA}

\history{Received on XXXXX; revised on XXXXX; accepted on XXXXX}
\editor{Associate Editor: XXXXXXX}
\maketitle

\begin{abstract}
\section{Summary:} BGT is a compact format, a fast command line tool and a
simple web application for efficient and convenient query of whole-genome
genotypes and frequencies across tens to hundreds of thousands of samples.
On real data, it encodes the haplotypes of 32,488 samples across 39.2
million SNPs into a 7.4GB database and decodes up to 420 million
genotypes per CPU second. The high performance enables real-time responses to
complex queries.

\section{Availability and implementation:} https://github.com/lh3/bgt

\section{Contact:} hengli@broadinstitute.org
\end{abstract}

\section{Introduction}

VCF/BCF~\citep{Danecek:2011qy} is the primary format for storing and analyzing
genotypes of multiple samples.  It however has a few issues. Firstly, VCF
is a site-oriented format. While accessing a site and all the associated
genotypes is efficient with indexing, retrieving
site annotations or the genotypes of a few samples always requires to decode
the genotypes of all samples, which is unnecessarily expensive. Secondly, VCF
does not take advantage of linkage disequilibrium (LD), while using this
information can dramatically improve compression ratio~\citep{Durbin:2014yq}.
Thirdly, a VCF record is not clearly defined. Each record may consist of
multiple alleles with each allele composed of multiple SNPs and INDELs. This
ambiguity complicates annotations, query of alleles and integration of multiple
data sets. At last, most existing VCF-based tools do not support expressive
data query. We frequently need to write scripts for advanced queries, which
costs both development and processing time. GQT~\citep{Ruan:2015ab} attempts
to solve some of these issues. While it is very fast for selecting a subset of
samples and for traversing all sites, it discards phasing, is inefficient for
region query and is not compressed well. The observations of these limitations
motivated us to develop BGT.

\begin{methods}
\section{Methods}
Unlike VCF which stores sample phenotypes, site annotations and genotypes all
in one file, BGT separates the three types of information into individual
files.  BGT keeps diploid genotypes as a 2-bit integer matrix $(H_{ki})$ with
row indexed by a pair of overlapping reference/non-reference alleles and column
by a sample haplotype (thus for $m'$ samples, the matrix has $2m'$ columns).
$H_{ki}$ takes value 0 if the $i$-th haplotype has the reference allele in the
allele pair at row $k$, takes 1 if the haplotype has the non-reference allele,
2 if unknown and 3 if the haplotype has a different non-reference allele. BGT
arbitrarily phases unphased genotypes and always breaks complex variants in VCF
down to the smallest possible variants. It keeps the allele pairs (i.e.  rows)
in a site-only BCF, disallowing multiple alleles per VCF line, and stores the
integer matrix as two positional BWTs (PBWTs), one for the lower bit and the
other for the higher bit.

BGT obtains phenotypes and site annotations from files in the Flat Metadata
Format (FMF). FMF is TAB-delimted with the first column showing the row name
and following columns giving typed key-value pairs. An example looks like:
\begin{center}
\begin{verbatim}
  sample1   gender:Z:M   height:f:1.73   foo:i:10
  sample2   gender:Z:F   height:f:1.64   bar:i:20
\end{verbatim}
\end{center}
BGT can retrieve rows via an arbitrary expression such as ``height$>$1.65''.

The multi-file design makes BGT unfriendly to data streaming, but it enables
BGT to use use one set of site annotations across multiple BGT files and allows
users to modify phenotypes and annotations without re-encoding all the genotypes.

\subsection{PBWT overview}

PBWT~\citep{Durbin:2014yq} is a generic way to encode binary matrices.  Let
$(A_k)_k=(A_0,\ldots,A_{n-1})$ denote a list of $m$-long binary strings. $(A_k)_k$
can be regarded as an $n\times m$ binary matrix with $A_k[i]$ representing the
cell at row $k$ and column $i$.  For simplicity, introduce a sentinel row
$A_{-1}=\$_0\$_1\cdots\$_{m-1}$ with a lexicographical order
$\$_0<\cdots<\$_{m-1}$.

Define binary string:
\[
P_{ki}=A_k[i]A_{k-1}[i]\ldots A_0[i]A_{-1}[i]
\]
to be the reverse of the \emph{column prefix} ending at $(k,i)$ and define $S_k(i)$ to
be the \emph{column index} of the $i$-th smallest prefix among set $\{P_{kj}\}_j$.
$S_k(i)$ is a bijection on $\{0,\ldots,m-1\}$ and thus invertible. In a
special case, $S_{-1}(i)=i$ because $P_{-1,i}=A_{-1}[i]=\$_i$.

The PBWT of $(A_k)_k$ is $(B_k)_k$, which is
calculated by
\begin{equation*}\label{eq:B}
B_k[i]=A_k[S_{k-1}(i)]
\end{equation*}
An important use of $(B_k)_k$ is to compute $S_k$. Define
\begin{equation*}\label{eq:phi}
\phi_k(i)=C_k(B_k[i])+{\rm rank}_k(B_k[i],i)
\end{equation*}
where $C_k(b)$ gives the number of symbols in $B_k$ that are lexicographically
smaller than $b$ and ${\rm rank}_k(b,i)$ the number of $b$ symbols in $B_k$
before position $i$. The $i$-th smallest column in row $k-1$ is ranked
$\phi_k(i)$ in row $k$. Thus
\begin{equation*}\label{eq:trans}
S_k(\phi_k(i))=S_{k-1}(i)
\end{equation*}
Given $A_k$ and $S_{k-1}$, we can compute $S_k$ and $B_k$ in the order of
$B_k\to\phi_k\to S_k$, starting from $k=0$. Conversely, given $B_k$ and
$S_{k-1}$, computing $\phi_k\to S_k\to A_k$ derives $A_k$ from $B_k$.

When there are strong correlations between adjacent rows, which is true for
haplotype data due to LD, $0$s and $1$s tend to form long
runs in $B_k$. This usually makes $B_k$ much more compressible than $A_k$ under
run-length encoding. For our test data set, 32 thousand genotypes in a row can
be compressed to less than 200 bytes in average.

\subsection{Query genotypes and output}

A BGT query may consist of three types of conditions: (a)
genotype-\emph{independent} sample selection, such as a list of sample names
or an arbitrary expression on phenotypes; (b) genotype-\emph{independent} site
selection, such as a genomic region, a list of alleles or an arbitrary
expression on variant annotations; (c) genotype-\emph{dependent} site
conditions, such as alleles being common among selected samples.
We may select multiple groups of samples with (a)-typed conditions. For each
group, BGT will compute aggregate variables, including the number of called
samples and the allele count, which can be outputted or used in (c)-typed
conditions.

BGT usually outputs VCF/BCF with aggregate variables written to the INFO field.
It may optionally output a TAB-delimited table on user selected fields. BGT
may also output the samples having a list of alleles, and the counts of
haplotypes across requested alleles in multiple sample groups.

%

\subsection{BGT server}

BGT comes with a standalone web server frontend implemented in the Go
programming language. The server has a similar interface to the command line
tool, but with additional consideration of sample anonymity. With BGT,
each sample has an attribute `minimal group size' or MGS. If a query selects a
group containing a sample with a MGS larger than the requested group size, the
server will refuse the request. In particular, if a sample has MGS larger than
one, users cannot access its sample name and individual genotypes, but can
retrieve allele counts computed together with other samples. This prevents
users to access data at the level of a single sample.

\end{methods}

\section{Results}

We generated the BGT database for the first release of Haplotype Reference
Consortium (HRC; http://bit.ly/HRC-org). The input is a BCF containing 32,488
samples across 39.2 million SNPs on autosomes. The BGT file size is 7.4GB, 11\%
of the genotype-only BCF, or 8\% of GQT. Decoding the genotypes of all samples
across 142k sites in a 10Mbp region takes 11 CPU seconds, which amounts to
decoding 420 million genotypes per second. This speed is even faster than
computing allele counts and outputting VCF.

We use the following command line to demonstrate the query syntax of BGT:
\begin{center}\footnotesize
\begin{verbatim}
  bgt view -G -d var.fmf.gz -a'gene=="BRCA1"' \
    -s 'source=="IBD"' -s 'source=="1000G"' \
    -f 'AC1/AN1>=0.001&&AC2/AN2>=0.001' \
   HRC-r1.bgt
\end{verbatim}
\end{center}
It finds BRCA1 variants annotated in `var.fmf.gz' that have $\ge$0.1\%
frequency in both the IBD data set (http://www.ibdresearch.co.uk) and
1000 Genomes~\citep{1000-Genomes-Project-Consortium:2012aa}. In this command line, {\tt -G} disables the output of genotypes.
Option {\tt -a} selects variants with the `gene' attribute equal to `BRCA1'
according to the variant database specified with {\tt -d}. This condition
is a (b)-typed condition independent of sample genotypes. Each option {\tt -s} sets an (a)-typed condition, selecting a group of
samples based on phenotypes.  For the \char35-th sample
group/{\tt -s}, BGT counts the total number of called alleles and the number of
non-reference alleles and writes them to the {\tt AN\char35} and {\tt
AC\char35} aggregate variables, respectively. Option {\tt -f} then use these
aggregate variables to filter output. This is a (c)-typed condition.

The command line above takes 12 CPU seconds with most of time spent on reading
through the variant annotation file to find matching alleles. The BGT server
reads the entire file into memory to alleviate the overhead, but a better
solution would be to use a proper database for variant annotations.

To demonstrate the server frontend, we have also set up a public BGT server at
http://bgtdemo.herokuapp.com. It hosts 1000 Genomes haplotypes for chromosome 11 and
20.

\section{Discussion}

Given a multi-sample VCF, most BGT functionalities can be achieved with small
scripts, but as a command line tool, BGT has a few advantages. Firstly, it
saves development time. Extracting information from multiple files can be done
with a command line instead of a script.  Secondly,
BGT saves processing time. With high-performance C code at the core, BGT is
much faster than processing VCF in a scripting language such as Perl or Python.
For example, deriving allele counts in a 10Mbp region for the HRC data takes 30
seconds with BGT, but doing the same with a Perl script takes 40 minutes, a
80-fold difference. Thirdly, the design of one non-reference allele per record
simplifies BGT merge and makes it twice as fast as bcftools merge on two
genotype-only input files.

The BGT server tries to solve a bigger problem: data sharing. Instead of always
delivering full data in VCF, projects could have a new option to serve data
publicly with the BGT server, letting users select the summary statistics of interest
on the fly while keeping samples unidentifiable. This is an improvement to
\citet{Stade:2014ty} which only provide precomputed summary.


We acknowledge that our MGS-based data sharing policy might have oversimplified
real scenarios, but we believe this direction, with proper improvements and
more importantly the approval of ethical review boards, will be more open,
convenient, efficient and secure than our current
share-everything-with-trust model.

\section*{Acknowledgement}
We are grateful to HRC for granting the permission to use the data for evaluating
the performance of BGT and thank the Global Alliance Data Working Group for the
helpful suggestions.
\paragraph{Funding\textcolon} NHGRI U54HG003037; NIH GM100233

\bibliography{bgt}

\end{document}